\documentclass[12pt]{article}

\usepackage{theorem,amssymb,amsbsy,latexsym}

\def\appendix#1{\addtocounter{section}{1}\setcounter{equation}{0}
\renewcommand{\thesection}{\Alph{section}}
\section*{Appendix \thesection\protect\indent \parbox[t]{11.715cm} {#1}}
\addcontentsline{toc}{section}{Appendix \thesection\ \ \ #1} }
\newcommand{\eq}{\begin{equation}}
\newcommand{\eqend}{\end{equation}}

\newcommand{\br}[1]{\left( #1 \right)}

\newcommand{\real}{{\bb R}} 

\font\mybb=msbm10 at 12pt
\def\bb#1{\hbox{\mybb#1}}

\def\nn{\nonumber}

\newcommand{\Tr}[1]{\:{\rm Tr}\,#1}
\def\e{{\,\rm e}\,}

\def\be{\begin{equation}}
\def\ee{\end{equation}}
\def\bea{\begin{eqnarray}}
\def\eea{\end{eqnarray}}
\def\bd{\begin{displaymath}}
\def\ed{\end{displaymath}}

\def\dd{{\rm d}}

\def\ii{{\,{\rm i}\,}}

\def\K{{{\rm K}_0}}
\def\K1{{{\rm K}_1}}

\makeatletter
\newdimen\normalarrayskip              
\newdimen\minarrayskip                 
\normalarrayskip\baselineskip
\minarrayskip\jot
\newif\ifold             \oldtrue            
\def\arraymode{\ifold\relax\else\displaystyle\fi} 
\def\@arrayskip{\ifold\baselineskip\z@\lineskip\z@
     \else
     \baselineskip\minarrayskip\lineskip2\minarrayskip\fi}
\def\@arrayclassz{\ifcase \@lastchclass \@acolampacol \or
\@ampacol \or \or \or \@addamp \or
   \@acolampacol \or \@firstampfalse \@acol \fi
\edef\@preamble{\@preamble
  \ifcase \@chnum
     \hfil$\relax\arraymode\@sharp$\hfil
     \or $\relax\arraymode\@sharp$\hfil
     \or \hfil$\relax\arraymode\@sharp$\fi}}
\def\@array[#1]#2{\setbox\@arstrutbox=\hbox{\vrule
     height\arraystretch \ht\strutbox
     depth\arraystretch \dp\strutbox
     width\z@}\@mkpream{#2}\edef\@preamble{\halign \noexpand\@halignto
\bgroup \tabskip\z@ \@arstrut \@preamble \tabskip\z@ \cr}%
\let\@startpbox\@@startpbox \let\@endpbox\@@endpbox
  \if #1t\vtop \else \if#1b\vbox \else \vcenter \fi\fi
  \bgroup \let\par\relax
  \let\@sharp##\let\protect\relax
  \@arrayskip\@preamble}
\makeatother

\newcommand{\beq}{\begin{eqnarray}}
\newcommand{\eeq}{\end{eqnarray}}

\begin{document}
\begin{titlepage}
\begin{flushright}

\baselineskip=12pt
HWM--03--6\\ EMPG--03--06\\ UUITP--02/03\\
hep-th/0303082\\
\hfill{ }\\ March 2003
\end{flushright}

\begin{center}

\baselineskip=24pt

{\Large\bf Exact Solution of Noncommutative Field Theory\\ in
  Background Magnetic Fields}

\baselineskip=14pt

\vspace{1cm}

{\bf E. Langmann}$^{1,}$\footnote{Email: {\tt
    langmann@theophys.kth.se}}, {\bf R.J. Szabo}$^{2,}$\footnote{Email:
  {\tt R.J.Szabo@ma.hw.ac.uk}} and {\bf
    K. Zarembo}$^{3,}$\footnote{Also at ITEP, Moscow, Russia. Email: {\tt
    Konstantin.Zarembo@teorfys.uu.se}}
\\[10mm]
$^1$ {\it Department of Physics -- Mathematical Physics, Royal Institute of
Technology\\ Stockholm Center for Physics, Astronomy and Biotechnology\\
SE-10691 Stockholm, Sweden}
\\[5mm]
$^2$ {\it Department of Mathematics, Heriot-Watt University\\ Scott
  Russell Building, Riccarton,
Edinburgh EH14 4AS, U.K.}
\\[5mm]
$^3$ {\it Department of Theoretical Physics, Uppsala University\\ Box
  803, SE-75108 Uppsala, Sweden}
\\[30mm]

\end{center}

\begin{abstract}

\baselineskip=12pt

We obtain the exact non-perturbative solution of a
scalar field theory defined on a space with noncommuting position and momentum
coordinates. The model describes non-locally interacting charged
particles in a background magnetic field. It is an exactly solvable
quantum field theory which has non-trivial interactions only when it
is defined with a finite ultraviolet cutoff. We propose that
small perturbations of this theory can produce solvable models with
renormalizable interactions.

\end{abstract}

\end{titlepage}

\newpage

\baselineskip=14pt

Quantum field theories on noncommutative spaces have received a surge
of interest in recent years, primarily because they can be obtained as
limits of string theory with background magnetic fields in which the
massive string modes decouple (see~\cite{ks} for reviews and
exhaustive lists of references). They capture many of the non-local
effects possessed by string theory but in a much simpler setting, and
have attained a fundamental level of interest as examples of non-local
field theories which may be well-defined. Various versions of them
have also been proposed as effective field theory descriptions of some
planar condensed matter systems in strong magnetic fields, such as
quantum Hall models. Because of their embedding into string theory,
these models are sometimes believed to be unitary and
renormalizable.\footnote{Disclaimer: Views and opinions mentioned in
  this letter do not necessarily reflect those of the authors.} However,
they possess several unusual aspects which continue to challenge the
conventional wisdom of quantum field theory, and question the
renormalizability and overall consistency of these field theories.

On a canonical noncommutative space, the usual pointwise product of fields
is replaced by the star-product
\beq
\Phi\star\Phi'(x)=\int\frac{\dd^Dk~\dd^Dq}{(2\pi)^D}~\tilde\Phi(k)\,
\tilde\Phi'(q)~\e^{\ii k_\mu\,\theta^{\mu\nu}\,q_\nu}~\e^{\ii(k+q)
\cdot x} \ ,
\label{starproddef}\eeq
where the tildes denote Fourier transforms and $\theta^{\mu\nu}$ is a
constant antisymmetric matrix.
The noncommutativity of space is encoded in the
fact that the commutators of coordinates computed with this
product are non-vanishing, as $x^\mu\star
x^\nu=x^\mu\,x^\nu+\ii\theta^{\mu\nu}$. In perturbation theory, the
phases in (\ref{starproddef}) produce momentum dependent vertices in
Feynman diagrams which affect the interactions of the quantum field
theory at energy scales below the scale $1/\sqrt{|\theta|}$ set by the
dimensionful noncommutativity parameter $\theta^{\mu\nu}$. The most
drastic example of this is known as
ultraviolet/infrared (UV/IR) mixing. If one uses Fourier expansion of
fields in a basis of plane waves $\e^{\ii p\cdot x}$, as in
(\ref{starproddef}), then the natural regularization of the quantum
field theory is the restriction of momenta $p$ to an annulus
$\Lambda_0<|p|<\Lambda$, where $\Lambda_0$ is an IR cutoff and
$\Lambda$ a UV cutoff. Removing the cutoffs amounts to
taking the limits $\Lambda_0\to0$ and $\Lambda\to\infty$. Planar
diagrams essentially coincide with those of ordinary quantum field
theory, while non-planar graphs are modified by phases containing
internal and external line momenta and are generically convergent. The
rapid phase oscillations in (\ref{starproddef}) imply that a
high-momentum cutoff $\Lambda$ generates an effective IR cutoff
$\Lambda_0=1/|\theta|\Lambda$. This appears to ruin the
usual Wilsonian renormalization procedure which would require a clear
separation of high and low momentum scales.

However, the puzzling UV/IR mixing properties may simply be an
artifact of perturbation theory which disappears when summed to all
orders. This is a non-perturbative issue which is in general
difficult to address. In this letter we will formulate a
noncommutative scalar field theory which is exactly solvable and obtain its
non-perturbative solution explicitly. The model describes charged
scalar particles in a background magnetic field with a four-point
interaction defined by the star-product~\cite{LS}. We will circumvent the
problems set in by UV/IR mixing by using a basis for the expansion of
fields on $\real^D$ which differs from the more conventional plane
wave basis and which will allow us to make sense of the field theory at a
fully non-perturbative level. This expansion provides a natural
non-perturbative regularization of the quantum field theory, producing
both a short-distance and low-momentum cutoff
simultaneously. We will show how to extract from this the exact
expressions for Green's functions of the quantum field theory.

The model is defined by the Euclidean action
\beq
S=\int\dd^2 x~\left[\Phi^*\left(H_B+m^2\right)\Phi+\frac g2\,
\Phi^*\star\Phi\star\Phi^*\star\Phi\right] \ ,
\label{Bmodelaction}\eeq
where $\Phi$ is a charged scalar field on flat space $\real^2$ and
\beq
H_B=\left(-\ii\partial_\mu-B\,\epsilon_{\mu\nu}\,x^\nu\right)^2
\label{LandauHam}\eeq
is the Landau Hamiltonian for a charged particle moving in two
dimensions under the influence of a constant perpendicularly applied
magnetic field $2B>0$. For brevity, we will only work in $D=2$
dimensions. The noncommutativity parameter is then given by
$\theta^{\mu\nu}=\theta\,\epsilon^{\mu\nu}$. Because the commutators
of the covariant momentum operators
$-\ii\partial_\mu-B\,\epsilon_{\mu\nu}\,x^\nu$ are equal to $-2\ii
B\,\epsilon_{\mu\nu}$, we may interpret the field theory
(\ref{Bmodelaction}) as being defined on a noncommutative space whose
corresponding momentum space is also given by noncommuting
coordinates. However, the ensuing analysis carries through to arbitrary even
dimensionality~\cite{langmann,inprep}, and remarkably most of our
conclusions hold quite generally. This follows from the fact that in
any even dimension $D$ there is a choice of coordinates which
skew-diagonalizes the problem into a product of $D/2$ two-dimensional
ones, to which the analysis of this letter applies with the
appropriate changes. The details will be presented in a separate
publication~\cite{inprep}.

We shall find that the UV fixed point of this theory is
trivial. The only scaling limit possible is one in which the coupling
constant $g$ vanishes as the UV cutoff is removed. We shall find that
there is no intermediate scale in between the two natural UV and IR
cutoffs in this model, consistent with the UV/IR
duality found in~\cite{LS}, and the field theory is not renormalizable
because the fields are correlated on the scale of the cutoff. This
result is similar in spirit to earlier observations that asymptotically-free
noncommutative field theories are trivial \cite{Akhmedov:2000uz}, and
that generic ones are only well-defined when they contain
both a finite UV and IR cutoff~\cite{amns}.
The renormalized propagator as an exact function of
external momentum is given in the scaling limit by
\beq
\tilde G(p)=\frac{\sqrt{(p^2+m^2)^2+4M^4}-(p^2+m^2)}{2M^4} \ ,
\label{Grenpexact}\eeq
where $m$ is the bare scalar particle mass and $M$ is a dynamically
generated mass scale. The non-free form of (\ref{Grenpexact}) takes
into account a non-perturbative resummation of leading power
divergences in the scaling limit which are generated by the
degeneracies of the Landau levels. Such a renormalization procedure,
though formally consistent, is physically  meaningless and has little chance to
 produce an interacting quantum field theory in the scaling limit. The
 scalar field theory is thereby an example of a noncommutative field
 theory, with a finite cutoff, which is exactly solvable. The exact
 propagator at finite cutoff, which is computed below, produces
 (\ref{Grenpexact}) in the scaling limit and has a qualitatively
 similar but somewhat more complicated form. It exhibits a novel
 oscillatory behaviour in position space on top of its long-distance
 exponential decay, which may be attributed to
 the appearence of an Aharonov-Bohm phase acquired by the charged
 particles in the magnetic background, similar to those observed
 numerically in~\cite{bhn}. Slight modifications of the model, such as
 the inclusion of a background harmonic oscillator potential that
 lifts the Landau level degeneracy, may produce good scaling limits.

Because of the magnetic field in the action,
a natural basis of normal modes
 is comprised of the orthonormal eigenfunctions
$\phi_{\ell,n}$ of the Landau Hamiltonian (\ref{LandauHam}),
\beq
H_B\phi_{\ell,n}=4B\left(\mbox{$\ell-\frac12$}\right)\,\phi_{\ell,n} \ ,
\label{Landaueigen}\eeq
with $\ell,n$ positive integers. Some properties of the Landau
eigenfunctions $\phi_{\ell,n}$ are briefly described in an appendix at
the end of this letter. These wavefunctions form the position space
representation of the occupation
number states $|\ell,n\rangle$ of two decoupled harmonic oscillators,
and with them we can expand the complex scalar fields of
(\ref{Bmodelaction}) as
\beq
\Phi(x)=\sqrt{4\pi\theta}~\mbox{$\sum_{\ell,n}$}\,A^{~}_{\ell n}~
\phi_{\ell,n}^*(x)
\label{Landauexp}\eeq
with $\phi_{\ell,n}^*=\phi_{n,\ell}^{~}$ and $A_{\ell n}$ dimensionless
complex numbers.

In this basis, the free part of the action (\ref{Bmodelaction}) is
diagonal, but the four-point star-product interaction term is
rather complicated. However, a special simplification occurs when the
parameters of the model are related through $B=1/\theta$.
In this case, the Landau wavefunctions have a remarkably simple
behaviour under star-products, $\phi_{\ell,n}\star\phi_{\ell',n'}=
\delta_{n\ell'}\,\phi_{\ell,n'}/\sqrt{4\pi\theta}$, which
can be derived by an explicit calculation and reflects the
fact that the one-particle wavefunctions
$\sqrt{4\pi\theta}\,\phi_{\ell,n}$ form the Wigner representations
of the Fock space operators $|\ell\rangle\langle n|$. As we show below,
the quantum field theory defined by (\ref{Bmodelaction}) is exactly
solvable precisely when the magnetic field and noncommutativity
parameter are related in this way, and we shall assume this
relation for the remainder of this letter. The action
(\ref{Bmodelaction}) then takes the simple form
\beq
S=\Tr\left[E\,A^\dag A+2\pi\theta g
\left(A^\dag A\right)^2\right] \ ,
\label{Bmodelmatrix}\eeq
where we have naturally assembled the expansion coefficients of
(\ref{Landauexp}) into an infinite complex matrix $A=(A_{\ell n})$ and
defined $E_{\ell n}=4\pi(4\ell-2+\theta m^2)\,\delta_{\ell n}$. The
noncommutativity of space is now manifested in the noncommutativity of
matrix multiplication in (\ref{Bmodelmatrix}).

This suggests that we may define the regularized quantum field
theory with action (\ref{Bmodelaction}) by restricting the quantum
numbers of the Landau wavefunctions to $\ell,n=1,\dots,N$ with
$N<\infty$. The path integral is then defined as the $N\to\infty$
limit of the $N\times N$ matrix integral
\beq
Z_N=\int\prod_{\ell,n=1}^N\,\dd A_{\ell n}^{~}~\dd A_{\ell
    n}^{*}~\e^{-\Tr\left[E\,A^\dag A+2\pi\theta g
\left(A^\dag A\right)^2\,\right]} \ .
\label{matrixintcalE}\eeq
The finite matrix dimension $N$ provides both a short-distance and
low-momentum cutoff simultaneously, because in matrix regularizations
of noncommutative field theory the UV and IR divergences are not
clearly separated and one needs to regulate them both at the same
time~\cite{LS,amns}. There are, however, many different ways to
take the large $N$ limit of the matrix model with partition function
(\ref{matrixintcalE}), and we need to decide which is the appropriate
one that captures the true non-perturbative physics of the original
continuum field theory.

The large $N$ limit is meaningful only when the entropy from the
growth in the number of integration variables is compensated by a
large action. In matrix models, an action of order $N^2$, typically of the form
$N~\Tr(\cdots)$, is necessary to balance quantum
fluctuations~\cite{'tHooft:1973jz}. Hence we need to require that
$\theta\sim N$ as $N\to\infty$ in (\ref{matrixintcalE}). In other
words, we must take the large $N$ limit while keeping fixed the ratio
\beq
\Lambda^2=N/4\pi\theta \ ,
\label{LambdaNtheta}\eeq
which simply defines the $\theta\to\infty$ limit of the noncommutative
field theory. This limit is a generic feature of the matrix
regularization of noncommutative field theories~\cite{amns,MRW}.
The important feature of the model with a background magnetic field is
that the whole action has a nice matrix representation, in contast to
other noncommutative field theories, in which the kinetic term has no
matrix representation~\cite{MRW}, and to ordinary field theories with
background fields, in which the kinetic term is simple in the Landau basis,
but the interactions are complicated~\cite{Kim:2001wb}.

There are two important consequences of this correlated large $N$ and
large $\theta$ limit. First of all,
this limit is just the standard 't~Hooft planar limit of the
matrix model. Secondly, the natural UV cutoff of the original
noncommutative field theory is the energy of the $N^{\rm th}$ Landau
level, which is $2B(2N-1)=16\pi\Lambda^2-2B$ and stays {\it finite}
as $N$ goes to infinity. Thus the quantity (\ref{LambdaNtheta}) is the
true UV cutoff of the quantum field theory. This will be confirmed
below by explicit calculations. Since $B\to0$ as
$N\to\infty$, the spacing between Landau levels also vanishes. Thus
taking the limit described above is equivalent to filling the finite
energy interval $[0,16\pi\Lambda^2]$ with infinitely many Landau
levels and an infinite density of states.

Thus the large $\theta$ limit of the model defined by
(\ref{Bmodelaction}) is a quantum field theory with a finite cutoff
whose exact solution is given by the 't~Hooft limit of the complex
external field matrix model (\ref{matrixintcalE}). As an example, we
will explicitly compute the exact two-point function defined by
\beq
G(x,y)=\Bigl\langle\Phi^*(x)\,\Phi(y)\Bigr\rangle=
4\pi\theta\,\sum_{\ell,n,\ell',n'}\,\Bigl\langle A_{\ell n}^*\,
A^{~}_{\ell'n'}\Bigr\rangle~\phi_{\ell,n}^{~}(x)\,\phi^{~}_{n',\ell'}(y) \ .
\label{propdef}\eeq
Both the action and integration measure in the path integral
(\ref{matrixintcalE}) are invariant under unitary transformations
$A\to U\cdot A$ with $U\in U(N)$.
This is just a consequence of the degeneracy of Landau levels.
We can make this transformation
explicit in the matrix integral and then integrate over the unitary
group. We then use the well-known properties of the Haar measure of
$U(N)$ and the fact that, by $U(N)$ invariance, the partition function
(\ref{matrixintcalE}) depends generically only on the $N$ eigenvalues
$\lambda_\ell$ of the external field $E/N$ and is symmetric under
permutation of them. It follows that the matrix averages appearing in
(\ref{propdef}) are given as $\langle
A_{\ell n}^*\,A^{~}_{\ell'n'}\rangle=-\frac1{N}\,\delta_{nn'}\,
\delta_{\ell\ell'}\,W(\lambda_\ell)$, where
\beq
W(\lambda_\ell)=\frac1N\,\frac{\partial\ln Z_N}{\partial\lambda_\ell}
\label{logderivZN}\eeq
and after differentiation the eigenvalues should be set equal to
$\lambda_\ell=16\pi\,\frac\ell N+\frac{m^2}{\Lambda^2}$. In what
follows it will prove convenient to shift
$\lambda_\ell\to\lambda_\ell-m^2/\Lambda^2$.

The computation of (\ref{propdef}) thereby boils down to the
calculation of the function (\ref{logderivZN}) and the sum over Landau
levels $\sum_n\phi_{\ell,n}(x)\,\phi_{n,\ell}(y)$
in the limit $N\to\infty$, $\ell\to\infty$ with
$\ell/N$ fixed. Using known properties of the Landau
wavefunctions, the sum can be evaluated in this scaling limit in terms
of the Bessel function $J_0$ of the first kind of order~0 (see the
appendix). In the large $N$ limit, we replace sums over Landau levels
by integrals in the standard way according to the rule
$\frac1N\,\sum_\ell\to\int_0^{16\pi}\dd\lambda/16\pi$, so that
(\ref{propdef}) becomes
\beq
G(x,y)=-\int_0^{16\pi}\frac{\dd\lambda}{4\pi}~W\left(\lambda+
\frac{m^2}{\Lambda^2}\right)\,J_0\Bigl(\Lambda\,\sqrt\lambda\,|x-y|\Bigr) \ .
\label{propJ0}\eeq
After a change of variables $\lambda=p^2/\Lambda^2$ and by using the
angular integral representation of the Bessel function, we can express
(\ref{propJ0}) as a two-dimensional integral
\beq
G(x,y)=-\frac1{\Lambda^2}\,\int_{|p|<4\,\sqrt\pi\,\Lambda}\,\frac{\dd^2p}
{(2\pi)^2}~W\left(\frac{p^2+m^2}{\Lambda^2}\right)~
\e^{\ii p\cdot(x-y)} \ .
\label{prop2Dint}\eeq
This result has several remarkable implications. First of all, it
demonstrates that the limit of large noncommutativity, in which the
underlying space is expected to degenerate and all
symmetries to be maximally violated, yields rotationally and
translationally invariant Green's functions. There are remnants of
UV/IR mixing in the far IR at $|x-y|\sim\sqrt\theta$, but these
distances have been scaled out and all results here are valid at
length scales far below the noncommutativity scale. Secondly, we see
that the quantity $4\,\sqrt\pi\,\Lambda$ is a sharp cutoff in the
momentum integral (\ref{prop2Dint}), showing clearly that
(\ref{LambdaNtheta}) is the UV cutoff of the field theory. Finally,
the matrix model partition function has the physical interpretation of
providing the exact propagator in momentum space through the function
(\ref{logderivZN}),
\beq
\tilde G(p)=-\frac1{\Lambda^2}\,W\left(\frac{p^2+m^2}{\Lambda^2}
\right) \ , ~~ p^2<16\pi\,\Lambda^2 \ .
\label{tildeGpmatrixfn}\eeq
It is instructive to consider a simple instance of this
identification. At zero coupling, the matrix integral
(\ref{matrixintcalE}) can be explicitly evaluated to $Z_N=\e^{-N\Tr\ln
  E}$, so that $W(\lambda)=-1/\lambda$. This recovers the expected
free propagator $\tilde G(p)=(p^2+m^2)^{-1}$.

It remains to compute (\ref{logderivZN}) in the general case.
This can be done rather explicitly, because this function satisfies in
the large $N$ limit a closed equation, which is the Schwinger-Dyson
equation of the matrix model given by
\beq
\frac{g}{\Lambda^2}\,\left(W^2(\xi)+\int_{m^2/\Lambda^2}^{16\pi+m^2/
\Lambda^2}\frac{\dd\lambda}{16\pi}
{}~\frac{W(\xi)-W(\lambda)}{\xi-\lambda}\right)
=\xi\,W(\xi)+1 \ .
\label{largeNloopeqn}\eeq
The loop equation (\ref{largeNloopeqn}) gives a straightforward way to
generate the perturbative expansion to arbitrary orders of the
original noncommutative field theory as an iterative solution of
(\ref{largeNloopeqn}) in the coupling constant $g$. By using
(\ref{tildeGpmatrixfn}), the propagator up to one-loop order is easily
determined in this way as
\beq
\tilde G(p)=\frac1{p^2+m^2}-\frac g{16\pi}\,\frac{\ln\left(16\pi\,
\Lambda^2/m^2\right)}{(p^2+m^2)^2}-\frac{g\,\Lambda^2}{(p^2+m^2)^3}+
O\left(g^2\right) \ .
\label{tildeGp1loop}\eeq
The second term in (\ref{tildeGp1loop}) recovers the usual one-loop
logarithmic UV divergence of $\Phi^4$ theory in two dimensions which
is generated by the planar (field theoretical) bubble diagram and
would lead to the renormalization group running of the mass. The third
term is an additional quadratic UV divergence which is the non-planar
(field theoretical) contribution. The additional divergences in
$\Lambda$ are even worse at higher loop orders. They arise from the
summations over degenerate Landau levels, whose degree of divergence grows
with the order of perturbation theory and differs from that of
usual scalar field theory.

The Schwinger-Dyson equation (\ref{largeNloopeqn}) can be solved by means
of the methods developed in \cite{Chekhov:1992hn} to give
\bea
W(\lambda)&=&\frac{\Lambda^2}{2g}\,
\left(\lambda-\sqrt{\lambda^2+a\,\lambda+b}\,\right)
\nn\\&&+\,\frac1{2}\,\int_{m^2/\Lambda^2}^{16\pi+m^2/\Lambda^2}
\frac{\dd\xi}{16\pi}~
\frac1{\sqrt{\xi^2+a\,\xi+b}}\,\frac{\sqrt{\lambda^2+a\,
\lambda+b}-\sqrt{\xi^2+a\,\xi+b}}{\lambda-\xi} \ .
\label{Wansatz}\eea
The parameters $a$ and $b$ are unambiguously determined by
substituting (\ref{Wansatz}) into (\ref{largeNloopeqn}), which
determines them through the algebraic equations
\bea
\int_{m^2/\Lambda^2}^{16\pi+m^2/\Lambda^2}
\frac{\dd\xi}{16\pi}~\frac1{\sqrt{\xi^2+a\,\xi+b}}
&=&\frac{a\,\Lambda^2}{2 g} \ , \label{abconstr1}\\
\int_{m^2/\Lambda^2}^{16\pi+m^2/\Lambda^2}
\frac{\dd\xi}{16\pi}~\frac{\xi}{\sqrt{\xi^2+a\,
\xi+b}}&=&\frac{\Lambda^2}{2 g}\,\left(b-\frac34\,a^2\right)-1 \ .
\label{abconstr2}\eea
The solution (\ref{Wansatz}) with these constraints matches the
perturbation expansion of (\ref{largeNloopeqn}) and has the correct
asymptotic behaviour $W(\lambda)\simeq-1/\lambda$ for
$\lambda\to\infty$. The loop amplitude $W(\lambda)$ is an analytic
function of $\lambda$ on the complex plane with a square-root branch
cut. The two branch points are the roots of the polynomial
$\lambda^2+a\,\lambda+b$ and are always complex, as follows from the
constraints (\ref{abconstr1}) and (\ref{abconstr2}).

{}From (\ref{prop2Dint}) it follows that the long-distance asymptotics
of the propagator are determined by the singularities of
$W(\lambda)$. Since the two branch points occur at complex $\lambda$,
the two-point function oscillates on top of its exponential
decay. Consequently, for $|x-y|\gg1/\Lambda$ we may write
$G(x,y)\simeq\e^{-|x-y|/L}$, where $1/L={\rm Im}\,p_0$ is determined
by the condition that $z=(p_0^2+m^2)/\Lambda^2$ solves the quadratic
equation $z^2+a\,z+b=0$. Careful inspection of the loop equations
shows that the correlation length $L$ is always of order of the cutoff scale
unless the coupling $g$ is very small, $g\sim1/\Lambda^2$,
and thus we define
\beq
g=M^4/\Lambda^2 \ .
\label{gMLambda}\eeq
{}From the constraint equations (\ref{abconstr1}) and (\ref{abconstr2})
we find that $b=2M^4/\Lambda^4$ and $a=O(M^4/\Lambda^4)$ in this scaling
limit. As a consequence, the renormalized two-point function,
which is the $\Lambda\rightarrow\infty$ limit of (\ref{tildeGpmatrixfn}),
reduces to (\ref{Grenpexact}). By examining (\ref{tildeGp1loop}) one
may infer that the scaling limit (\ref{gMLambda}) resums the leading power
divergences arising in the perturbation series. The explicit form of
the propagator (\ref{tildeGpmatrixfn}) at finite cutoff $\Lambda$ is given by
(\ref{Wansatz})--(\ref{abconstr2}).

The power divergences arising in perturbation theory spoil the
renormalizability of this field theory, but there can be many ways to get rid
of them. For instance, we can replace the Landau Hamiltonian
(\ref{LandauHam}) in (\ref{Bmodelaction}) by the combination
$H_B+\sigma H_{-B}$, with $\sigma$ a small parameter. Physically,
this corresponds to the addition of a confining electric potential to
the background of the charged scalar fields. This extension lifts the
degeneracy of the Landau levels, yet the regulated version of the field theory
 still reduces to the matrix model (\ref{matrixintcalE}) with an additional
term $\sigma\Tr\,E\,AA^\dag$ in the action. While this term spoils the
$U(N)$ invariance of the matrix model, the latter still has a regular
large $N$ limit and is potentially solvable by an extension of the techniques
presented in this letter by perturbative expansion in $\sigma$. The
special case $\sigma=1$ corresponds to charged particles in a harmonic
oscillator potential alone and is closest to the conventional
noncommutative field theories with no background magnetic field.

\noindent{\bf Acknowledgments:} We thank G.~Akemann, J.~Ambj\o rn,
H.~Braden, D.~Johnston and V.~Kazakov for helpful
discussions. E.L. and K.Z. would like to thank the Erwin-Schr\"odinger
Institute in Vienna for hospitality during part of this work. This
work was supported in part by the Swedish Science Research Council~(VR)
and the G\"oran Gustafssons Foundation. The work of R.J.S. was supported in
part by an Advanced Fellowship from the Particle Physics and Astronomy Research
Council~(U.K.). The work of K.Z. was supported in part by
RFBR grant 01--01--00549 and grant 00--15--96557 for the promotion of
scientific schools.

\bigskip

\noindent{\bf Appendix:} Here we collect some pertinent properties of Landau
eigenfunctions. By introducing two sets of creation and
annihilation operators $a=\partial/\sqrt{B}+\sqrt{B}\,\bar{z}/2$,
$a^\dagger=-\bar{\partial}/\sqrt{B}+\sqrt{B}\,z/2$
and $b=\bar{\partial}/\sqrt{B}+\sqrt{B}\,z/2$,
$b^\dagger=-\partial/\sqrt{B}+\sqrt{B}\,\bar{z}/2$, with $z=x^1+\ii x^2$
and $\partial= (\partial_1-\ii\partial_2)/2$,
 the Landau Hamiltonian (\ref{LandauHam}) can be written as
 $H_B=4B\,\left(a^\dagger\,a+\mbox{$\frac12$}\right)$. The
 eigenfunctions $\phi_{\ell,n}(z,\bar z)$ of this Hamiltonian are
characterized by the occupation numbers associated with the $a$ and
$b$ oscillators and can be conveniently written in terms of the
generating function (see~\cite{langmann}, for example)
\begin{equation}
F_{s,t}(z,\bar{z})\equiv\sum_{\ell,n=1}^\infty \frac{s^{\ell-1}\,t^{n-1}}
{\sqrt{(\ell-1) !\,(n-1)!}}~\phi_{\ell,n}(z,\bar{z})
=\sqrt{\frac{B}{\pi}}~\e^{-B\,|z|^2/2+\sqrt{B}\,(sz+t\bar{z})-st} \ .
\end{equation}
A straightforward calculation of the star-product with $\theta=1/B$
yields the identity $F_{s,t}\star
F_{s',t'}=\e^{s't}~F_{s,t'}/\sqrt{4\pi\theta}$, from which the formula
for the star-product of Landau wavefunctions used in the main text may
be easily deduced.

The sum over Landau levels that was encountered in the calculation
of the two-point function can be found as follows.
To compute $g_\ell(x,y)=4\pi\theta\,\sum_n\phi_{\ell,n}(x)\,
\phi_{n,\ell}(y)$, we introduce the generating function
 $g(x,y;r)=\sum_\ell\,g_\ell(x,y)\,\frac{r^{2\ell}}{\ell!}$. This can be
calculated as
\begin{eqnarray}
g(x,y;r)&=&4\pi\theta\,\int\frac{\dd^2u}{\pi}~\e^{-|u|^2}~\int_0^{2\pi}
\frac{\dd\varphi}{2\pi}~
F_{r\e^{\ii\varphi},u}(x)\,F_{\bar{u},r\e^{-\ii\varphi}}(y)
\nonumber \\
&=&4~\e^{-\frac{1}{2\theta}\,|x-y|^2+\frac{\ii}{\theta}\,x\times y+r^2}
{}~J_0\br{2r\,|x-y|/\sqrt{\theta}\,},
\label{gxyr}\end{eqnarray}
where $x\times y=\epsilon_{\mu\nu}\,x^\mu\,y^\nu$. By extracting the
Taylor coefficients of (\ref{gxyr}) using contour integration, we
get in the limit of large $\ell$ and large $\theta$ the result
$g_\ell(x,y)=4\,J_0(2\,|x-y|\sqrt{\ell/\theta}\,)$ that was used in the
main text.

\end{document}